\documentclass[12pt]{article}

\usepackage{scicite}

\usepackage{times}
\usepackage{amsmath,amssymb}

\usepackage{graphicx}
\usepackage[colorlinks=true
,urlcolor=blue
,anchorcolor=blue
,citecolor=blue
,filecolor=blue
,linkcolor=red
,menucolor=blue
,linktocpage=true
,pdfproducer=medialab
,pdfa=true
]{hyperref}

\newenvironment{sciabstract}{%
\begin{quote} \bf}
{\end{quote}}

\title{Response to a comment on Dessert et al.
``The dark matter interpretation of the 3.5 keV line is
inconsistent with blank-sky observations"}

\author
{Christopher Dessert,$^{1}$ Nicholas L. Rodd,$^{2,3}$ Benjamin R. Safdi$^{1}$\\
\\
\normalsize{$^{1}$Leinweber Center for Theoretical Physics, Department of Physics,}\\
\normalsize{University of Michigan, Ann Arbor, MI 48109, USA}\\
\normalsize{$^{2}$Berkeley Center for Theoretical Physics,}\\
\normalsize{University of California, Berkeley, CA 94720, USA}\\
\normalsize{$^{3}$Theoretical Physics Group, Lawrence Berkeley National Laboratory,}\\
\normalsize{Berkeley, CA 94720, USA}
}

\date{}

\begin{document} 


\baselineskip24pt


\maketitle


\begin{sciabstract}
The dark matter explanation of the 3.5 keV line is strongly disfavored by our work in~\cite{Dessert:2018qih}.
Ref.~\cite{Boyarsky:2020hqb} questions that conclusion: modeling additional background lines is claimed to weaken the limit sufficiently to re-allow a dark matter interpretation.
We respond as follows.
1) A more conservative limit is obtained by modeling additional lines; this point appeared in its entirety in our work in~\cite{Dessert:2018qih}, though we also showed that the inclusion of such lines is not necessary.
2) Despite suggestions in~\cite{Boyarsky:2020hqb}, even the more conservative limits strongly disfavor a decaying dark matter origin of the 3.5 keV line.
\end{sciabstract}

In~\cite{Dessert:2018qih} (DRS20) we presented evidence disfavoring a dark matter origin for the 3.5 keV line~\cite{Bulbul:2014sua}.
We did so by exploiting the full archival XMM-Newton dataset, and the insight that every observation includes a column density of the Milky Way dark-matter halo.
The absence of a line at 3.5 keV in this dataset excluded the best fit dark-matter parameter space by roughly two orders of magnitude.
The authors of~\cite{Boyarsky:2020hqb} (BRMS) have claimed our limits are overstated by an order of magnitude, and once corrected resurrect the dark-matter interpretation.
We disagree for reasons given below.

We first consider their central argument that ``the dark matter interpretation of the 3.5 keV signal remains viable" after the proposed modifications to the fiducial DRS20 analysis.
BRMS contend that our fiducial limit should be replaced with our most conservative results, which appeared in Fig. S14(A) and accounted for possible astrophysical or instrumental lines at 3.3 keV and 3.68 keV.
Over the sterile neutrino mass range $m_s \in [6.95, 7.15]$ keV, relevant to explain an unidentified X-ray line at 3.5 keV, the weakest fiducial 95\% limit in DRS20 on the mixing angle is $\sin^2(2\theta) \lesssim 2 \times 10^{-12}$, whereas the more conservative value that arose from modeling the additional lines was $\lesssim 10^{-11}$.
BRMS claimed that the conservative limit revives the dark-matter hypothesis.
We will now show it does not, even with a generous accounting of statistical and astrophysical uncertainties.
To do so we will discuss previous Galactic and extragalactic observations of the line (illustrated in Fig. 1 of DRS20).

The most direct comparison is to a detection with Chandra of the 3.5 keV line in the Milky Way, as claimed in~\cite{Cappelluti:2017ywp} (5 in Fig. 1 of DRS20).
DRS20 derived limits using observations between $5^\circ$-$45^\circ$ of the Galactic Center (GC), whereas~\cite{Cappelluti:2017ywp} detected the line at $\sim$115$^\circ$.
Thus a comparison only requires an accounting for the shape of the Milky Way halo (the normalization, {\it e.g.} the local dark-matter density, is common to both datasets).
The dark-matter interpretation of the line observed in~\cite{Cappelluti:2017ywp} implies a best-fit mixing angle of central value of $\sin^2(2\theta) \approx 1.9 \times 10^{-10}$ (as low as $4.8 \times 10^{-11}$ allowing for a 2$\sigma$ downward statistical fluctuation, though note that the detection in~\cite{Cappelluti:2017ywp} was only $\sim$3$\sigma$ in local significance) and using their assumed Navarro-Frenk-White (NFW) dark-matter profile.
Even the mixing angle found when assuming a 2$\sigma$ downward fluctuation is in strong tension with our conservative limit ($p$-value $< 10^{-15}$ ).
If we instead assume a Burkert dark-matter profile with a 9 kpc core radius, the dark-matter densities at $\sim$115$^\circ$ and $5^\circ$-$45^\circ$ are enhanced and reduced, respectively.
Assuming this profile together with a $2$$\sigma$ downward statistical fluctuation in~\cite{Cappelluti:2017ywp}, the preferred mixing angle in~\cite{Cappelluti:2017ywp} may be at minimum  $3.0 \times 10^{-11}$, while the conservative 95\% limit in DRS20 becomes $\lesssim 1.3 \times 10^{-11}$, again incompatible ($p < 10^{-5}$).
The 2$\sigma$ lower value for the mixing angle in~\cite{Cappelluti:2017ywp} is still excluded at 95\% confidence by the conservative limit in DRS20 even if one assumes the two regions have the same $D$-factors (the dark-matter density integrated along the line-of-sight, $D = \int ds\,\rho_{\rm}(s)$).

The 3.5 keV line was observed in stacked clusters by the MOS and PN cameras onboard {XMM-Newton}~\cite{Bulbul:2014sua} (3 and 4 in Fig. 1 of DRS20).
Again allowing for 2$\sigma$ downward statistical fluctuations the mixing angle may be as low as $\sin^2(2\theta) \approx 3.7 \times 10^{-11}$ ($3.3 \times 10^{-11}$) for MOS (PN) with best-fit values $6.0 \times 10^{-11}$ ($5.4 \times 10^{-11}$), where we exclude the Perseus, Coma, Centaurus, and Ophiuchus results, as these preferred larger and even more excluded mixing angles ({\it e.g.}, the mixing angle needed to explain their Perseus MOS result is $\sin^2(2\theta) =  55_{-16}^{+26} \times 10^{-11}$, while  Coma + Centaurus + Ophiuchus prefer  $\sin^2(2\theta) = 18^{+4}_{-4} \times 10^{-11}$).
 We have generously allowed for the larger $D$-factors presented in~\cite{Bulbul:2014sua}.
These are $\sim$50\% higher on average than more modern estimates, which have well-quantified and uncorrelated uncertainties $\sim$20-40\% per cluster~\cite{Lisanti:2017qlb}.
Even with the more optimistic $D$-factors and allowing for 2$\sigma$ downward statistical fluctuations these mixing angles remain in strong tension with a conservative interpretation of DRS20, $\sin^2(2\theta) \lesssim 2 \times 10^{-11}$.
To obtain this conservative limit, we start from Fig. S14(A) of DRS20 and then assume a Burkert profile and a local dark-matter density of 0.24 GeV/cm$^3$, which is 2$\sigma$ below the value preferred by a recent {\it Gaia} rotation-curve analysis~\cite{Eilers}.
To emphasize the robustness of this limit we could even push to an extreme scenario  where the dark-matter profile is a constant-density sphere within the inner 21 kpc with total mass within that radius as measured in~\cite{21kpc} using Globular Clusters; this would only weaken the limit to $\sin^2(2\theta) \lesssim 2.8 \times 10^{-11}$.

Ref.~\cite{Boyarsky:2014jta} detected the 3.5 keV line in XMM-Newton data from M31 (2 in Fig. 1 of DRS20), with central value $\sin^2(2\theta) \approx 6 \times 10^{-11}$ (note that Ref.~\cite{Boyarsky:2014jta} incorrectly converts flux to mixing angle for M31).
Allowing for a 2$\sigma$ downward fluctuation and a $\sim$70\% increase in the M31 $D$-factor (the largest value suggested in~\cite{Boyarsky:2014jta}), the mixing angle could be as low as $1.8 \times 10^{-11}$, approximately equal to the conservative limits in DRS20 only after extremely conservative assumptions on both.
Further, more modern studies prefer a smaller $D$-factor for M31.
Ref.~\cite{Boyarsky:2014jta} used a central value of $D = 600$ $M_\odot/{\rm pc}^2$, whereas a comprehensive analysis in \cite{Tamm:2012hw} preferred at most $D = 300 \pm 60$ $M_\odot/{\rm pc}^3$, in which case the dark-matter line in M31 would be even more strongly excluded.
The smaller $D$-factor is consistent with other recent catalogs~\cite{Lisanti:2017qlb}.
The most conservative limit in DRS20 clearly strongly excludes the decaying dark-matter interpretation of the 3.5 keV line, but as discussed in DRS20 our fiducial limit is even stronger and well justified; it passed a litany of consistency checks and no evidence for the necessity of additional lines on top of the existing background models and within the 0.5 keV energy window was found.

Next, we address BRMS' claim that considering a 1 keV width energy window (as opposed to the 0.5 keV window used in DRS20) results in a weaker limit and a 4$\sigma$ detection of a line at 3.48 keV.
This conclusion was reached using a simplified version of the DRS20 analysis in a partially overlapping dataset.
In detail they perform a stacked analysis with a single power-law background, as well as potential additional lines.
DRS20 performed a joint likelihood analysis, where each exposure is modeled separately with independent power-law models for the astrophysical and QPB data.
DRS20 selected low-background exposures, while BRMS did not; BRMS use 17 Ms of MOS data between $20^{\circ}$--$35^{\circ}$ of the GC, while only 8.5 Ms passes the DRS20 cuts.

\begin{figure}[!t]
\includegraphics[width = 0.99\textwidth]{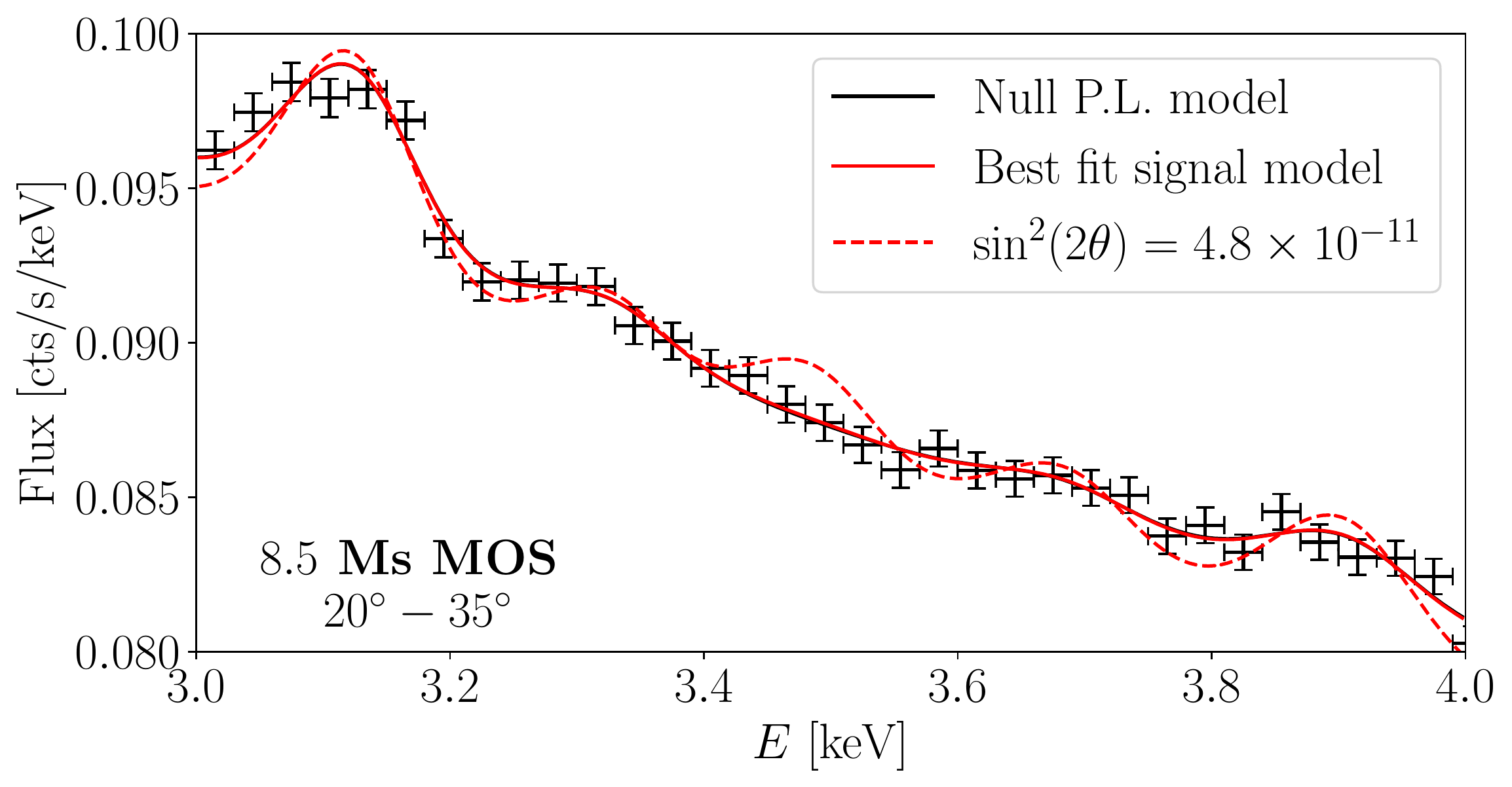}
\vspace{-0.4cm}
\caption{\noindent 
Our rendition of BRMS' analysis for the subset of their data that passes the quality cuts in DRS20.  While BRMS claimed a 4$\sigma$ signal at 3.48 keV, we find no evidence for a line in the vicinity of 3.48 keV from the fit over the energy range 3 - 4 keV.  We illustrate the fit with a fixed signal at the 2$\sigma$ lowest value allowed to explain the Chandra deep-field observation~\cite{Cappelluti:2017ywp}.  Data in this plot is down-binned by a factor of 6 solely for presentation, and this analysis is publicly available~\cite{jupyter}. Note the null model is almost indistinguishable from the best fit signal model.
}
\label{fig:main}
\end{figure}

In Fig.~\ref{fig:main} we repeat the analysis of BRMS on the 8.5 Ms of data actually used in DRS20.
The flux is a factor of $\sim$2 lower than in BRMS, highlighting the importance of quality cuts.
When modeling, we follow BRMS precisely and include a power-law background model and lines at 3.12, 3.32, 3.68, and 3.90 keV with floated normalizations, in addition to a line at 3.5 keV for the signal hypothesis.
We show the best-fit null and signal models; the $\Delta \chi^2$ between the two is $\sim$0.03 and the reduced $\chi^2$ value for the null hypothesis fit is $\chi^2 / {\rm DOF} \approx 190/194$, indicating that the null model is describing the data to the level of statistical noise.
The one-sided 95\% upper limit on the 3.5 keV line from this analysis, using the fiducial dark-matter model from DRS20, is $\sin^2(2\theta) \lesssim 10^{-11}$ (or in flux $0.015~{\rm cts/cm^2/s/sr}$).
We further illustrate the best fit model with fixed $\sin^2(2\theta) = 4.8\times 10^{-11}$, the lowest allowed value (at 2$\sigma$) in the Chandra analysis~\cite{Cappelluti:2017ywp}; that model is disfavored relative to the null by $\Delta \chi^2 \approx 40$.
The analyses and datasets we used to reach these conclusions are provided in supplementary \texttt{Jupyter} notebooks~\cite{jupyter}, where we also show that this result is robust to choices of background model, energy range, and data-selection criterion, and further that we can qualitatively (though not quantitatively) reproduce BRMS' results on their dataset (but we show that their claimed evidence for a 3.48 keV line even in that lower-quality dataset is not robust~\cite{jupyter}).

In summary, the decaying dark-matter origin of the 3.5 keV line is strongly constrained.

\section*{Additional Comments}

Here we address the issue raised in BRMS regarding the out-of-time event subtraction. BRMS disagree with a choice we made in the data reduction process for the PN data. Our choice returns an integer counts array, as required for our analysis, and was recommended to us by the XMM Science team. The alternatively suggested choice returns non-integer counts. Our choice increases the background rate by an energy-independent 4\%, and as such has an entirely neglibible effect on our results. 

\section*{Acknowledgments}

This work benefited from the feedback of Tesla Jeltema, Mariangela Lisanti, Siddharth Mishra-Sharma, Kerstin Perez, Stefano Profumo, and Tracy Slatyer.
CD and BRS are supported by the Department of Energy Early Career Grant DE-SC0019225.
NLR is supported by the Miller Institute for Basic Research in Science at the University of California, Berkeley.

\end{document}